\documentclass[10pt,epsfig]{article}
\usepackage{graphicx}
\usepackage{amssymb}

\title{The Topological Nature Of Defects}
\author{ E.D.M. Kavoussanaki\thanks{E-mail: kele@physics.uoc.gr} \\
University of Crete, Department of Physics, \\
P.O. Box 2208, 71303 Heraklion, Crete, Greece}

\begin{document}

\maketitle


\begin{abstract}
The subject of topological defects has become a very attractive
field of study given its apparent relevance to as diverse systems
as the early universe and condensed matter. As usually envisaged
the topology of the manifold $M$ of the minima of the relevant to
our physical system effective potential $V$, provides reliable
information about the capacity of that system to accommodate
topological defects. Here we will examine the premises for that
statement to be true.
\end{abstract}

\section{Introduction}

An \textit{ordered medium} corresponds to some real smooth
manifold $W$ where a function $\varphi : W\rightarrow M$ is
defined which assigns to every point of the medium an order
parameter. For $x$ a point of $W$, the order parameter corresponds
to the value of $\varphi (x)$. In our discussion, in fact, we will
refer to $\varphi$ itself by the name \textit{order parameter}
even though this title might be confusing since one could think in
terms of a map from the ordered medium to a space of functions,
allocating a function (the order parameter) at every point of the
manifold. In our discussion, we will call $\varphi$ by what it
actually does: it designates a value, the order parameter, at
every point of $W$.

Excluding the trivial case where the order parameter is constant
throughout the medium (which is called, thus, uniform), we will
focus our attention to non uniform media where the function,
through connected space, varies continuously apart perhaps
(depending on the specific configuration) at isolated
regions\cite{mermin}.

The ordered medium has no defects if the order parameter is
everywhere continuous. The defect will be associated with a space
region containing the discontinuity of the order parameter. For
example, assume that the order parameter $\varphi
:W=\mathbb{R}\rightarrow M=\mathbb{R}$ is given by
\[
\varphi (x)=\left\{
\begin{array}{ll}
f(x) & x>x_{0} \\
c & x=x_{0} \\
f(x) & x<x_{0}
\end{array}
\right.
\]
for some constant $c$ and $f:\mathbb{R}\rightarrow\mathbb{R}$ .
For example, $f(x)$ could be
$\frac{(c_1\Theta(x-x_0)+c_2\Theta(x_0-x))}{(1+\Theta(x-x_0)\Theta(x_0-x))}$
with $\Theta(x)=1$ for $x\ge 0$ and zero otherwise. Take $f(x)$ to
be a continuous function for all $x\ne x_0$. Thus, we can
calculate the
\[
\lim_{x\rightarrow x_{0}^{\pm }}f(x)
\]
which we will call as $f^{+}(x_{0})$ and $f^{-}(x_{0})$
accordingly. If $f^{+}(x_{0})\ne f^{-}(x_{0})$ then, regardless of
the actual value of $c=\varphi (x_{0})$, $\varphi$ will be
discontinuous at $x=x_{0}$. Otherwise, for $f ^{+}(x_{0})=f
^{-}(x_{0})$ then, if there is a discontinuity of $\varphi$ at
$x=x_{0}$, it will be due to the value of $c$ which should satisfy
the relation: $c\ne \bigg[f^{+}(x_0)=f^{-}(x_0)\bigg]$. Both cases
assimilate a topological defect (at $x=x_0$) since $\varphi$, for
different reasons for each situation, is discontinuous at $x_0$.
For the record, we will associate the first type of discontinuity
with a stable defect and the second with an unstable one.

Correlating discontinuities in $\varphi$ with topological defects,
is a quite controversial way forward. The main reason is that
one does not need discontinuities to establish whether or not
topological defects exist within a certain field configuration.
However, here we will advocate in favor of the argument
maintaining that we can always appropriately associate a
topological defect with a discontinuity, the latter being the
primordial source of defects.

Consider a function $\Psi :W\rightarrow \mathcal{M}$ to be
continuous and to incorporate defects. We should be able to
recover a discontinuity on the defects. This can be done by
substituting $\Psi : W\rightarrow \mathcal{M}$ with a
$\varphi:W\rightarrow M$ where $M \subset \mathcal{M}$, and
$\varphi=\Psi$ away from the defects. At the latter, $\varphi$ is
discontinuous.

The converse is not necessarily true. That is, suppose we know
$\Psi : W\rightarrow \mathcal{M}$ to be a continuous function but
we have no information about the existence of topological defects
in its configuration. If we can find an $M\subset\mathcal{M}$ and
define a $\varphi:W\rightarrow M$ so that $\varphi$ to be
discontinuous on $W$, then we cannot automatically draw an analogy
between the points of discontinuity in $\varphi$ with possible
embodied topological defects in $\Psi$.

If we know of the existence of topological defects, we can either
directly or indirectly recover discontinuities at the defects.
However, discontinuities by themselves do not necessarily indicate
the presence of topological defects. For the latter to be securely
identified one needs more information than simply the reassurance
of the feasibility of singularities for some appropriate choice of
field $\varphi$. There are energetic conditions one should verify
whether or not they are satisfied.

Most physical systems are described by a continuous function. The
ordered medium corresponds to some physical system which is being
described by a continuous function $\Psi$. For classical physical
systems, at least, one assumes a tendency for the system to
acquire the lowest possible energy state. Since our discussion
will be mainly on a classical level, we will adopt the viewpoint
that this tendency for the lowest energy state is incorporated in
the behavior of $\Psi$. The lowest energy state we will call the
``ideal" or ``perfect" state which $\Psi$ aims for.

If some $\Psi$ does not correspond to the lowest energy we will
say that it must incorporate defects, being faithful perhaps to
the true meaning of the word, since such a configuration will have
to be ``imperfect". The energy of some $\Psi$ will be measured in
terms of the Hamiltonian energy density
$\mathcal{H}=\mathcal{K}+\mathcal{V}$ with $\mathcal{K}$ and
$\mathcal{V}$ the kinetic and potential energy density
contributions respectively. We are going to postulate here that
the occurrence of any type of topological defect will be a
consequence of the departure from the ground state of
$\mathcal{H}$. If this is some inevitable result of certain
conditions then the arising defect will be called stable.
Otherwise, it will be identified as an unstable defect.

When $\Psi$ corresponds to the lowest energy state, it should be a
continuous map from $W$ to $\mathcal{M}$, where $\mathcal{M}$ the
manifold of the minima of the potential energy $\mathcal{V}$. In
that case, $\Psi$ does not incorporate defects. When $\Psi$ does
not correspond to the lowest energy, topological defects must be
there. We are going to postulate that the topological defects
within $\Psi$ exist as discontinuities in $\varphi$. In fact, a
defect corresponds to an appropriate change in $\varphi$, applied
only to those regions where it is discontinuous, having as a main
aim to restore the continuity of that function throughout the
original medium. Thus, the discontinuities will indicate where the
introduction of a defect is needed. Associating, however, those
configurations with a particular order parameter behavior is
another matter which we will look into later.

It is convenient to call $M$ the manifold of the minima of the
potential energy density $\mathcal{V}$ and assume that $\varphi$,
should be a function from the ordered medium $W$ to $M$. Ideally,
$\varphi : W\rightarrow M$ ought to be a continuous function on
$W$ and equal $\Psi$, describing the lowest energy state. However,
the minimal energy state cannot always be achieved. We will
suppose that this will be due to the existence of discontinuities
in $\varphi$ and therefore assimilated topological defects in
$\Psi$\footnote{In case $\Psi : W\rightarrow \mathcal{M}$
incorporates defects then $\mathcal{M}$ cannot be the manifold of
the minima of $\mathcal{V}$. However, it should be possible to
define a function $\varphi : W\rightarrow M$ with $M$ the manifold
of the minima of $\mathcal{V}$, which will be discontinuous at the
defects.}. Therefore, the ground state cannot be reached because
$\varphi$ cannot be a continuous function to $M$.

Our contemplation to consider order parameters with
discontinuities might seem quite far fetched to someone having in
mind the situation in physical systems where the order parameter
is always continuous. We shall see, though, that thinking about
discontinuities is quite interesting and leads to intriguing
results. First, we are going to investigate whether there is a
pattern in the occurrence of discontinuities for some order
parameter $\varphi$, physical space $W$ and minimal energy
manifold $M$.

\section{Considering discontinuities}

In our discussion here we will focus on stable defects for these
are the configurations that can be uniquely associated with
certain topological characteristics of specific manifolds.
Unstable defects, on the other hand, are heavily dependent on the
definition of $\varphi$ and they are not necessarily derived from
the topological properties of the manifolds $W$ or $M$. We shall
postpone a discussion about them for later.

Consider the manifold of the minima to be $M=\{-1,1\}$, or any
discrete set, and $W$ to be any simply connected manifold of
dimension $D\ge 1$. A function $\varphi : W\rightarrow M$ will
necessarily be discontinuous on $W$ when any two points $a$ and
$b$ on $W$ have $\varphi (a)\ne \varphi (b)$. 
The stable defect will happen if we try to associate two points on
$W$ that {\it can} drop to one another on $W$ with two points of
$M$ that {\it cannot} drop to one another on $M$. The arising
stable defects are identified as \textit{domain walls}.

Take $M=S^{1}$ and $W$ either $\mathbb{R}^{2}$ or
$\mathbb{R}^{3}$. It is clear that any closed contour on $W$ can
shrink continuously to a point. This is in contrast to what
happens on $M$ where no closed contour\footnote{Of the form
$g:I\equiv[0,1]\subset\mathbb{R}\rightarrow M$ such that
$g(0)=g(1)$ and $g(t_1)\ne g(t_2)$ for $t_1\ne t_2$ for each
$t_1$, $t_2$ $\epsilon (0,1)$} can contract to a point while
remaining a closed contour. The stable defect will happen if we
try to associate a loop and a point on $W$ that {\it can} turn
into one another on $W$ with a loop and a point on $M$ that {\it
cannot} share that property on $M$.
Those kind of stable defects are called \textit{cosmic strings}.

Consider $M=S^{2}$ and $W$ to be $\mathbb{R}^{w}$ with $w>2$. As
before, one links a $2$-sphere and a point in $W$ that {\it can}
transform into one another on $W$, with a $2$-sphere and a point
on $M$ that {\it cannot} do that. The associated stable
topological defects are called \textit{magnetic monopoles}.

In general, consider two configurations on $W$ that can
continuously drop to one another. Those configurations are
homotopic on $W$. Define a function $\varphi$ on them so that
their corresponding configurations on $M$ will not share their
homotopy property. Inevitably, $\varphi$ will incorporate at least
one stable defect. Going a step ahead, the postulation we are
advocating for here is that the stable defects are a consequence
of the forced association of configurations that belong to
different homotopy classes of a certain homotopy group. That
association happens because of the difference in the topological
properties of $W$ and $M$ and of the way $\varphi$ is defined on
$W$.

\section{Homotopy Groups}

Our postulation here is that stable topological defects happen
when one tries to link two different values of a quantized
quantity. In particular, we argue that the discontinuity arises
because we have to link two maps that belong to different homotopy
classes of $\pi_k(M)$ for some $k$. The quantized quantity
therefore corresponds to the Hopf invariant or winding number that
turns out only to depend on the homotopy class a map to $M$
belongs to.

We can start by recalling some basic notions of the theory of
homotopy groups of spheres\cite{baez}. There are often lots of
topologically different ways of wrapping an $k-$dimensional sphere
around a $m-$dimensional sphere. The group of all homotopy classes
of ways of wrapping an $k$-sphere around an $m$-sphere is the
$\pi_k(S^m)$. Only the maps which belong to the same homotopy
class can be deformed into one another continuously. The different
homotopy classes are classified by an integer, the ``winding
number" which indicates how many times the $k$-sphere has been
wrapped around the $m$-sphere.

The task of calculating the $\pi_k(S^{m})$ for all $k$ and $m$ is
not at all simple. Regardless of the actual form of
$\pi_k(S^{m})$, the reassurance that it is non trivial can be
enough to conclude that topological defects are possible. Since we
need to associate two maps that belong to different homotopy
classes of $\pi_k(M)$ in order to create a topological defect, it
is clear that $\pi_k(M)$ must accommodate different homotopy
classes and thus be non trivial. There are some results we need to
keep in mind. The homotopy group is trivial, thus
$\pi_k(S^{m})=\{0\}$, when $k<m$. Further,
$\pi_k(S^{k})=\mathbb{Z}$ for all $k\ge 1$ and
$\pi_k(S^{1})=\{0\}$ for all $k>1$.

For example, consider maps from the unit circle in the complex
plane to itself. Two functions $g_0(\theta)=1$ and
$g_1(\theta)=e^{\imath 2\pi\cos^{2}\frac{\theta}{2}}$ with
$\theta$ the polar angle, are homotopic since one can define a
homotopy $G$ as $G(t,\theta)=e^{\imath
2t\pi\cos^{2}\frac{\theta}{2}}$ so that $g_0$ becomes $g_1$ and
vice versa. That is $G(0,\theta)=g_0$ and $G(1,\theta)=g_1$ since
$t\epsilon I\equiv[0,1]$. In fact, any continuous function from
the (complex unitary) circle to itself can be continuously
deformed to exactly one of the functions $f(z)=z^{n}$ with $n$ the
winding number, an integer, and $\parallel z\parallel =1$. We know
the homotopy class of a map from a circle to itself if we know its
winding number.

Let $f:S^{k}\hookrightarrow \mathbb{R}^{k+1}$ be the inclusion of
the $k$-sphere. Call
\begin{equation}F:I\otimes S^{k}\rightarrow
\mathbb{R}^{k+1}\label{Fdef}\end{equation} a homotopy that drops
$f$ to a point of $\mathbb{R}^{k+1}$. Due to convexity of
$\mathbb{R}^{k+1}$, we can write $F$ as
\begin{equation}F(t,z)=F_t(z)=(1-t)f(z)+tc\label{Fhomo}\end{equation}
with $c$ a point on $\mathbb{R}^{k+1}$ and $z$ a point on the
$k$-sphere.

Suppose $h:S^{k}\rightarrow S^{m}$ a map from a $k$-sphere to an
$m$-sphere. That map should belong to a non trivial class of
$\pi_k(S^{m})\ne\{0\}$ where $k\ge m$. Let
\begin{equation}H:I\otimes S^{k}\rightarrow
S^{m}\label{Hdef}\end{equation} be a homotopy written as
$H(q,z)=H_q(z)$ such that $H(0,z)=h(z)$ and $H(1,z)=h_f(z)$ with
$h_f$ another function of the homotopy class $h$ belongs to and
$z$ a point on the $k$-sphere.

{\sc Question}: Is there a way to make $q$ a function of $t$ so as
to allow \begin{equation} G: ImF_t\rightarrow
ImH_q\label{Ghomo}\end{equation} to be a continuous well defined
onto function for every $t$?

The reason for requiring $G$ to be surjective is that we would not
like to give $G$ the freedom of choosing its range. This can
restrict the way $G$ is defined over its domain. For each $0\le
t<1$, $ImF_t$ corresponds to a $k$-sphere in $\mathbb{R}^{k+1}$
and we can define $G$ as \begin{equation}G\equiv H\circ F:
I\otimes I\otimes S^{k}\rightarrow
S^{m}\label{Hcombo}\end{equation} For the considered $t$ interval,
$q$ can be any smooth function of $t$ and the construction in
(\ref{Hcombo}) is well defined. However, when $t=1$, $F_1$ changes
to dimension zero and $H\circ F_1$ is always a point on the
$S^{m}$. This means that $G$ cannot be onto $ImH_{q(1)}$ with the
latter being the area spanned by $H(q(1),z)$ for every $z$ on a
$k$-sphere. If we assume $q(1)=1$, we can say that there is no
$H(1,z)=h_f(z)$ that will permit $G$ to be well defined on $F_1$.

On the other hand, if $H$ is a homotopy between functions of the
trivial homotopy class of $\pi_k(S^{m})$, we can require
$H(1,z)=c_0$, a constant, and $h_i(z)=H(0,z)$ some null homotopic
function. We can define $G$ as in (\ref{Hcombo}) and rest assured
that so long as $q(t)$ satisfies the equation $q(1)=1$, we can
always find an $h_i(z)$ so that $G$ is well defined.

Stable topological defects will occur when $G$ cannot be well
defined for a particular combination of $F$ and $H$ homotopies.

\section{The occurrence of stable topological defects}

Consider $\varphi$ the order parameter defined on $W$. The $ImF_t$
with $F$ as defined in (\ref{Fdef}) must be homeomorphic to a
subset of $W$ for each $t\epsilon I$. We can suppose $W$ to be a
convex subset of $\mathbb{R}^{w}$ with $w\ge k+1$. Assume we know
$\varphi_0=\varphi(F_0(z))$ and $\varphi_1=\varphi(F_1(z))$ with
$F_1(z)$ a point on $W$ and $ImF_0$ homeomorphic to an $S^{k}$.
Let $\varphi\circ F_0:S^{k}\rightarrow M$ belong to some homotopy
class $[a]$ of $\pi_k(M)\ne \{0\}$. The $\varphi\circ
F_1:S^{k}\rightarrow M$ will necessarily belong to the trivial
class of the $k$-homotopy group of $M$. The latter is considered
some $m$-sphere. We must have $1\le m\le k\le w-1$.

Call $H:I\otimes S^{k}\rightarrow M=S^{m}$ a homotopy between
$\varphi:S^{k}\subset W\rightarrow M$ and some
$\varphi_f:S^{k}\subset W\rightarrow M$. The stable topological
defect will occur when we cannot find a continuous function
$H(q(t),F(t,z))$ so that
$H(q(0),F(0,z))=\varphi_0=\varphi(F(0,z))$ and
$H(q(1),F(1,z))=\varphi_1=\varphi(F(1,z))$ with $q(t)$ a
continuous function of $t\epsilon I$ with $q(1)=1$ and $q(0)=0$.
In the absence of defects, $H(q(t),F(t,z))$ may be considered to
be the order parameter.

To be sure, the discontinuity in $G$, when stable topological
defects exist, will be evident as $t$ changes. That is, if we do
not keep track of the changes in $F$ and $G$ as $t$ changes then
we will not be able to detect the singularity. Consider, for
example, $\varphi: W=S^1\rightarrow M=S^1$ as
\[\varphi(\theta)=e^{\imath\theta\exp(-\sin^{2}(\frac{\theta}{2}))}\] with
$\theta$ the polar angle on $W$ and $\theta\epsilon[0,2\pi]$. Call
$\vartheta=\theta\exp(-\sin^{2}(\frac{\theta}{2}))$ One could
introduce another variable, say $t$, which would alter the
definition of $\varphi$ on $W$ as
\begin{equation}
e^{\imath(\vartheta\circ\vartheta^{\Theta(t-t_1)}\circ\vartheta^{\Theta(t-t_2)}\circ\ldots
)} \label{texture}
\end{equation}
with increasing $t\epsilon\mathbb{R}^{+}$ (e.g $t_2=t_1+\Delta
t_{21}$ with $\Delta t_{21}$ a certain step and $t_1>0$). The
function $\Theta$ corresponds to the Theta function which, here,
is defined as $\Theta(x\ge 0)=1$ and $\Theta(x<0)=0$. We assume
that $\vartheta^0=1$ and $\vartheta$ is always finite. We see that
the function in (\ref{texture}), as $t\rightarrow\infty$, tends to
behave in a similar fashion as the $\Theta(\theta-2\pi)$ with
$\theta\epsilon[0,2\pi]$. At every individual $t$, the function on
$W$ would be given by (\ref{texture}) and it would be continuous
on $W$. However, if we were to draw the changes the order
parameter sustained as $t$ changed we would discover
discontinuities. The discontinuities would be visible on the space
$\mathbb{R}^{+}\otimes S^{1}$ and not the $S^{1}$ alone.

\section{Conclusions}
The usual premise for the feasibility of stable topological
defects is the $\pi_k(M)\ne\{0\}$. If $\pi_k(W)\ne\{0\}$ then it
is not necessary that $\pi_k(M)\ne\{0\}$ can provide reliable
information on the occurrence of topological defects. Consider,
for example, $M=S^{m}=W=S^{w}$. We must have $k\le w$ if we want
to find an inclusion map of a $k$-sphere to $W$. Further, $k\ge
m\ge 1$ since $\pi_k(M)\ne\{0\}$. Therefore, $k=w$. In that
situation $\pi_k(S^{w})=\pi_k(S^{k})=\mathbb{Z}$. The homotopy
$F(t,z)$ may relate maps of a non trivial homotopy class of
$\pi_k(W)$. If that happens then $\varphi\circ F(0,z)$ and
$\varphi\circ F(1,z)$ have no reason to belong to different
homotopy classes of $\pi_k(M)$. In fact, stable topological
defects cannot happen in that situation.

Consider $\Phi_0=\varphi\circ F(0,z)$ a function to a non trivial
homotopy class of $\pi_k(M)$. The $\varphi\circ F(1,z)$ will
necessarily belong to a trivial class, because $F(1,z)$ is a
single point on $W$. That is, we can write $\varphi\circ F(1,z)$
as $\Phi_1=\tilde{\varphi}\circ F(0,z)$ with $\tilde{\varphi}$ a
constant function. We should not be able to find a homotopy
between $\Phi_0$ and $\Phi_1$. That is the cause of the stable
defect. In other words, the stable defect arises not only because
$\Phi_0$ belongs to a non trivial class but also because $F(1,z)$
becomes such a configuration on $W$ that $\varphi$, when defined
on $F(1,z)$ will {\it have to} create a trivial map $\Phi_1$.

Usually, one assumes that $M$ is the boundary of $W=D^{w}$, a disk
or otherwise a convex subset of $R^{w}$. Thus, $W=D^{w}$.
Therefore, one fixes $M$ to some $S^{w-1}$ and therefore the only
way that $\pi_k(S^{w-1})$ is non trivial and the $k$-sphere could
contract to point in $W$, is if $k=w-1$. To be sure,
$\pi_k(S^{m})$ can be non trivial for various values of $k$. The
ordered medium as much as the dimension $m$, will dictate the
values of $k$ that can be relevant. The general framework we
presented here aims at providing the means to creating various
situations where defects should arise.

This can be achieved by considering, for instance, different forms
for the $F$ homotopy in (\ref{Fdef}). We can also examine what may
happen for different $W$s. For example, assume that $W$ is a
$2$-sphere and $m=1$. Therefore, $1\le k\le 1\Rightarrow k=1$. We
see that there are two degenerate functions $F$ as in (\ref{Fdef})
that can realize the homotopy to a point on $W$. There are, thus,
two points of discontinuity and two stable topological defects
arising as a result. The study of such situations will be reported
elsewhere.

\section{Acknowledgments}
The author would like to thank Dr. R.J. Rivers,
Professor T.W.B. Kibble and Professor S. Papadopoulou for useful
discussions.

\end{document}